\documentclass[11pt]{article}
\usepackage{epsfig}
\newcommand{\ct}{\tt}
\newcommand{\cf}{\texttt}

\begin{document}

\title{Flavor: A Language for Media Representation}
\author{Alexandros Eleftheriadis, Danny Hong, and Yuntai Kyong \\
        Department of Electrical Engineering\thanks
                 {This material is based upon work supported in part by
                  the National Science Foundation under Grant MIPS-9703163.} \\
        Columbia University \\
        New York, NY 10027, USA
}
\date{January 6, 2003}
\maketitle

\begin{abstract}
Flavor (Formal Language for Audio-Visual Object Representation) has been 
created as a language for describing coded multimedia bitstreams in a formal
way so that the code for reading and writing bitstreams can be automatically
generated.  It is an extension of C++ and Java, in which the typing system 
incorporates bitstream representation semantics. This allows describing in a 
single place both the in-memory representation of data as well as their 
bitstream-level (compressed) representation. Flavor also comes with a 
translator that automatically generates standard C++ or Java code from the
Flavor source code so that direct access to compressed multimedia information 
by application developers can be achieved with essentially zero programming. 
Flavor has gone through many enhancements and this paper fully describes the 
latest version of the language and the translator. The software has been made 
into an open source project as of Version 4.1, and the latest downloadable 
Flavor package is available at {\ct http://flavor.sourceforge.net}.
\end{abstract}

\section{Introduction} 
\label{sec:intro}

Flavor originated from the need to simplify and speed up the development of software 
that processes coded audio-visual or general multimedia information.  This includes 
encoders and decoders as well as applications that manipulate such information. 
Examples include editing tools, synthetic content creation tools, multimedia indexing 
and search engines, etc.  Such information is invariably encoded in a highly efficient 
form to minimize the cost of storage and transmission.  This source coding~\cite{bk:cover:info} 
operation is almost always performed in a bitstream-oriented fashion: the data to be 
represented is converted to a sequence of binary values of arbitrary (and typically
variable) lengths, according to a specified syntax.  The syntax itself can have
various degrees of sophistication.  One of the simplest forms is the GIF87a
format~\cite{mn:gif}, consisting of essentially two headers and blocks of coded 
image data using the Lempel-Ziv-Welch (LZW) compression.  Much more complex formats include 
JPEG~\cite{mn:jpeg}, MPEG-1~\cite{mn:mpeg1}, MPEG-2~\cite{mn:mpeg2,bk:haskell:mpeg2} 
and MPEG-4~\cite{mn:mpeg4,pr:signal:imcomm00}, among others.

General-purpose programming languages such as C++~\cite{bk:stroustrup:c++} and Java~\cite{bk:arnold:java} 
do not provide native facilities for coping with such data.  Software codec 
(encoder/decoder) or application developers need to build their own facilities, 
involving two components.  First, they need to develop software that deals with the 
bitstream-oriented nature of the data, as general-purpose microprocessors are strictly 
byte-oriented.  Second, they need to implement parsing and generation code that 
complies with the syntax of the format at hand (be it proprietary or standard).  These 
two tasks represent a significant amount of the overall development effort.  They also 
have to be duplicated by everyone who requires access to a particular compressed 
representation within their application.  Furthermore, they can also represent a 
substantial percentage of the overall execution time of the application.

Flavor addresses these problems in an integrated way.  First, it allows the ``formal'' 
description of the bitstream syntax.  Formal here means that the description is based on 
a well-defined grammar, and as a result is amenable to software tool manipulation.  In 
the past, such descriptions were using ad-hoc conventions involving tabular data or 
pseudo-code.  A second and key aspect of Flavor's architecture is that this description 
has been designed as an extension of C++ and Java, both heavily used object-oriented 
languages in multimedia applications development.  This ensures seamless integration of 
Flavor code with both C++ and Java code and the overall architecture of an application.

Flavor was designed as an object-oriented language, anticipating an audio-visual world 
comprised of audio-visual objects, both synthetic and natural, and combining it with 
well-established paradigms for software design and implementation.  Its object-oriented 
facilities go beyond the mere duplication of C++ and Java features, and introduce several 
new concepts that are pertinent for bitstream-based media representation.
 
In order to validate the expressive power of the language, several existing bitstream 
formats have already been described in Flavor, including sophisticated structures such as 
MPEG-2 Systems, Video and Audio.  A translator has also been developed for translating 
Flavor code to C++ or Java code.  We should note that Flavor is currently used in the 
Structured Audio as well as the Systems parts of the MPEG-4 standard.

Emerging multimedia representation techniques can directly use Flavor to represent the 
bitstream syntax in their specifications.  This will allow immediate use of such 
specifications in new or existing applications, since the code to access/generate 
conforming data can be generated directly from the specification with zero cost.  In 
addition, such code can be automatically optimized; this is particularly important for 
operations such as Huffman decoding/encoding, a very common tool in media representation.

In the following, we first present a brief background of the language in terms of its 
history and technical approach.  We then describe each of its features, including 
declarations and constants, expressions and statements, classes, scoping rules, maps, and 
built-in operators.  We also describe the translator and its simple run-time API.  Finally, 
we conclude with an overview of the benefits of using the Flavor approach for media 
representation.  More detailed information and publicly available software can be found in 
the Flavor web site at: {\ct http://flavor.sourceforge.net}.  Parts of this paper have been 
presented in~\cite{ipr:fang:flavor,ipr:eleft:flavor,ipr:fang:map}.

\section{Background} 
\label{sec:background}

\subsection{A Brief History}
\label{sec:history}

Flavor has its origins in a Perl script ({\ct mkvlc})~\cite{mn:mkvlc} that was
developed in early 1994 in order to automate the (laborious) generation of C
code declarations for variable-length code (VLC) tables of the MPEG-2 Video
specification\footnote{The {\ct mkvlc} program is available at {\ct
http://www.ee.columbia.edu/$\sim$eleft/software/mkvlc.tar.gz}.}.  In November 1995,
the ideas behind {\ct mkvlc} took a more concrete shape in the form of a
``syntactic description language,''~\cite{ipr:fang:flavor,mn:eleft:sdl} i.e., 
a formal way to describe not just VLCs, but the entire structure of a bitstream. 
Such a facility was proposed to the MPEG-4 standardization activity, which at 
that time had started to consider flexible, even programmable, audio-visual 
decoding systems.  The language subsequently underwent a series of revisions 
obtaining input from several participants in the MPEG-4 standardization 
activity, and its specification is now fairly stable.

\subsection{Technical Approach}
\label{sec:tech}

Flavor provides a formal way to specify how data is laid out in a serialized 
bitstream.  It is based on a {\it principal of separation} between bitstream 
parsing operations and encoding, decoding and other operations.  This separation
acknowledges the fact that the same syntax can be utilized by different tools, 
but also that the same tool can work unchanged with a different bitstream 
syntax.  For example, the number of bits used for a specific field can change
without modifying any part of the application program.

Past approaches for syntax description utilized a combination of tabular data,
pseudo-code, and textual description to describe the format at hand.  Taking
MPEG as an example, both MPEG-1 and MPEG-2 specifications were described using
a C-like pseudo-code syntax (originally introduced by Milt Anderson,
Bellcore), coupled with explanatory text and tabular data.  Several of the
lower and most sophisticated layers (e.g., macroblock) could only be
handled by explanatory text.  The text had to be carefully crafted and tested
over time for ambiguities.  Other specifications (e.g., GIF and JPEG) use
similar bitstream representation schemes, and hence share the same
limitations.

Other formal facilities already exist for representing syntax.  One example is 
ASN.1 (ISO International Standards 8824 and 8825).  A key difference, however, 
is that ASN.1 was not designed to address the intricacies of source coding 
operations, and hence cannot cope with, for example, variable-length coding. 
In addition, ASN.1 tries to hide the bitsream representation from the developer
by using its own set of binary encoding rules, whereas in our case the binary 
encoding is the actual target of description.

There is also some remote relationship between syntax description and 
``marshalling,'' a fundamental operation in distributed systems where 
consistent exchange of typed data is ensured.  Examples in this category include
Sun's ONC XDR (External Data Representation) and the {\ct rpcgen} compiler 
which automatically generates marshalling code, as well as CORBA IDL, among 
others.  These ensure, for example, that even if the native representation of an
integer in two systems is different (big versus little endian), they can still 
exchange typed data in a consistent way.  Marshalling, however, does not 
constitute bitstream syntax description because: 1) the programmer does not 
have control over the data representation (the binary representation for each 
data type is predefined), 2) it is only concerned with the representation of 
simple serial structures (lists of arguments to functions, etc.).  As in ASN.1, 
the binary representation is ``hidden'' and is not amenable to customization by
the developer.  One could parallel Flavor and marshalling by considering the 
Flavor source as the XDR layer.  A better parallelism would be to view Flavor as
a parser-generator like {\ct yacc} \cite{bk:levine:lex}, but for bitstream 
representations.

It is interesting to note that all prior approaches to syntactic description
were concerned only with the definition of message structures typically found
in communication systems. These tend to have a much simpler structure compared
with coded representations of audio-visual information (compare the UDP packet
header with the baseline JPEG specification, for example).

A new language, Bitstream Syntax Description Language (BSDL)~\cite{mn:bsdl1,mn:bsdl2},
has recently been introduced in MPEG-21~\cite{mn:mpeg21} for describing the 
structure of a bitstream using XML Schema.  However, unlike Flavor, BSDL is 
developed to address only the high-level structure of the bitstream, and it
becomes almost impossible to fully describe the bitstream syntax on a 
bit-per-bit bases.  For example, BSDL does not have a facility to cope with 
variable-length coding, whereas in Flavor, map (described in Section~\ref{sec:maps})
can be used.  Also, the BSDL description would be overly verbose, requiring 
a significant effect to review and modify the description with human eye.

Flavor was designed to be an intuitive and natural extension of the typing 
system of object-oriented languages like C++ and Java.  This means that the 
bitstream representation information is placed together with the data 
declarations in a single place.  In C++ and Java, this place is where a class 
is defined.

Flavor has been explicitly designed to follow a declarative approach to 
bitstream syntax specification.  In other words, the designer is specifying how 
the data is laid out on the bitstream, and does not detail a step-by-step 
procedure that parses it.  This latter procedural approach would severely limit
both the expressive power as well as the capability for automated processing 
and optimization, as it would eliminate the necessary level of abstraction.  As
a result of this declarative approach, Flavor does not have functions or 
methods.

A related example from traditional programming is the handling of floating 
point numbers.  The programmer does not have to specify how such numbers are 
represented or how operations are performed; these tasks are automatically 
taken care of by the compiler in coordination with the underlying hardware or 
run-time emulation libraries.

An additional feature of combining type declaration and bitstream 
representation is that the underlying object hierarchy of the base programming
language (C++ or Java), becomes quite naturally the object hierarchy for 
bitstream representation purposes as well.  This is an important benefit for 
ease of application development, and it also allows Flavor to have a very rich
typing system itself.
 
``{\bf HelloBits}'' - 
Traditionally, programming languages are introduced via a simple ``Hello
World!'' program, which just prints out this simple message on the user's
terminal.  We will use a similar example with Flavor, but here we are concerned
about bits, rather than text characters.  Figure~\ref{fig:hellobits} shows a
set of trivial examples indicating how the integration of type and bitstream
representation information is accomplished.  Consider a simple object called
{\ct HelloBits} with just a single value, represented using 8 bits.  Using the
MPEG-1/2 methodology, this would be described as shown in Figure~\ref{fig:hellobits}(a). 
A C++ description of this single-value object would include two methods to 
read and write its value, and have a form similar to the one shown in Figure~\ref{fig:hellobits}(b). 
Here {\ct getuint()} is assumed to be a function that reads bits from 
the bitstream (here 8) and returns them as an unsigned integer (the most 
significant bit first); the {\ct putuint()} function has similar functionality
but for output purposes.  When {\ct HelloBits::get()} is called, the bitstream 
is read and the resultant quantity is placed in the data member {\ct Bits}. 
The same description in Flavor is shown in Figure~\ref{fig:hellobits}(c).

As we can see, in Flavor the bitstream representation is integrated with the type 
declaration.  The Flavor description should be read as: {\ct Bits} is an
unsigned integer quantity represented using 8 bits in the bitstream.  Note that 
there is no implicit encoding rule as in ASN.1: the rule here is embedded in
the type declaration and indicates that, when the system has to parse a
{\ct HelloBits} data type, it will just read the next 8 bits as an unsigned
integer and assign them to the variable {\ct Bits}.

\begin{figure}[htp!]
\begin{center}

\begin{tabular}{c}

\begin{tabular}{|lrl|}                                            
\hline
Syntax                 & No. of bits      & Mnemonic \\ \hline
\multicolumn{3}{|l|}{\cf{HelloBits} \{}              \\
\hspace{.2in}\cf{Bits} & 8                & uimsbf   \\
\}                     &                  &          \\ 
\hline
\end{tabular} \\
(a) \\
\ \\
\begin{tabular}{|l|} 
\hline
\cf{class HelloBits} \{                \\
\hspace{.2in}\cf{unsigned int Bits};   \\
\hspace{.2in}\cf{void get()} \{        \\
\hspace{.4in}\cf{Bits = ::getuint(8)}; \\
\hspace{.2in}\}                        \\
\hspace{.2in}\cf{void put()} \{        \\
\hspace{.4in}\cf{::putuint(8, Bits)}; \\
\hspace{.2in}\}                        \\
\};                                    \\ 
\hline
\end{tabular} \\
(b) \\
\ \\
\begin{tabular}{|l|}
\hline
\cf{class HelloBits} \{                 \\ 
\hspace{.2in}\cf{unsigned int(8) Bits}; \\
\};                                     \\ 
\hline
\end{tabular} \\
(c)

\end{tabular}

\caption{\cf{HelloBits}. 
         (a) Representation using the MPEG-1/2 methodology.
         (b) Representation using C++ (A similiar construct would be used for Java as well).
         (c) Representation using Flavor.}
\label{fig:hellobits}
\end{center}
\end{figure}

These examples, although trivial, demonstrate the differences between the 
various approaches.  In Figure~\ref{fig:hellobits}(a), we just have a
tabulation of the various bitstream entities, grouped into syntactic units. 
This style is sufficient for straightforward representations, but fails when 
more complex structures are used (e.g., VLCs).  In Figure~\ref{fig:hellobits}(b), 
the syntax is incorporated into hand-written code embedded in a {\ct get()} 
and {\ct put()} or an equivalent set of methods.  As a result, the syntax 
becomes an integral part of the encoding/decoding method even though the same 
encoding/decoding mechanism could be applied to a large variety of similar 
syntactic constructs.  Also, it quickly becomes overly verbose.

Flavor provides a wide range of facilities to define sophisticated bitstreams,
including {\ct if-else}, {\ct switch}, {\ct for}, and {\ct while} constructs.
In contrast with regular C++ or Java, these are all included in the data
declaration part of the class, so they are completely disassociated from code
that belongs to class methods.  This is in line with the declarative nature of
Flavor, where the focus is on defining the structure of the data, not
operations on them.  As we show later on, a translator can automatically
generate C++ and/or Java methods ({\ct put()} and {\ct get()}) that can read
or write data that complies to the Flavor-described representation.

In the following we describe each of the language features in more detail, 
emphasizing the differences between C++ and Java.  In order to ensure that 
Flavor semantics are in line with  both C++ and Java, whenever there was a 
conflict a common denominator approach was used.

\section{Language Overview}
\label{sec:lang}

\subsection{Declarations and Constants}
\label{sec:decl_and_const}

\subsubsection{Literals}
\label{sec:literals}

All traditional C++ and Java literals are supported by Flavor.  This includes
integers, floating-point numbers and character constants (e.g., {\ct `a'}). Strings
are also supported by Flavor.  They are converted to arrays, with or without a
trailing `\verb+\0+' (null character).

Additionally, Flavor defines a special binary number notation using the prefix
{\ct 0b}. Numbers represented with such notation are called binary literals
(or bit strings) and, in addition to the actual value, also convey their
length.  For example, one can write {\ct 0b011} to denote the number 3
represented using 3 bits. For readability, a bit string can include periods
every four digits, e.g., {\ct 0b0010.01}.  Hexadecimal or octal constants used
in the context of a bit string also convey their length in addition to their
value.  Whenever the length of a binary literal is irrelevant, it is treated as
a regular integer literal.

\subsubsection{Comments}
\label{sec:comments}

Both multi-line \verb+(/**/)+ and single-line \verb+(//)+ comments are allowed.
The multi-line comment delimiters cannot be nested.

\subsubsection{Names}
\label{sec:names}

Variable names follow the C++ and Java conventions (e.g., variable names cannot
start with a number).  The keywords that are used in C++ and Java are considered
reserved in Flavor.

\subsubsection{Types}
\label{sec:types}

Flavor supports the common subset of C++ and Java built-in or fundamental
types.  This includes {\ct char}, {\ct int}, {\ct float}, and {\ct double}
along with all appropriate modifiers ({\ct short}, {\ct long}, {\ct signed}
and {\ct unsigned}).  Additionally, Flavor defines a new type called {\ct bit}
and a set of new modifiers, {\ct big} and {\ct little}.  The type {\ct bit} is
used to accommodate bit string variables and the new modifiers are used to
indicate the endianess of bytes.  The {\ct big} modifier is used to represent
the numbers using big-endian byte ordering (the most significant byte first)
and the {\ct little} modifier is used for the numbers represented using the
little-endian method.  By default, big-endian byte ordering is assumed.  Note
that endianess here refers to the bitstream representation, not the processor
on which Flavor software may be running.  The latter is irrelevant for the
bitstream description.

Flavor also allows declaration of new types in the form of classes (refer to
Section~\ref{sec:classes} for more information regarding classes).  However,
Flavor does not support pointers, references, casts, or C++ operators related
to pointers.  Structures or enumerations are not supported either, since they
are not supported by Java.

\subsubsection{Declarations}
\label{sec:decls}

Regular variable declarations can be used in Flavor in the same way as in C++ 
and Java.  As Flavor follows a declarative approach, constant variable 
declarations with specified values are allowed everywhere (there is no 
constructor to set the initial values).  This means that the declaration 
`{\ct const int a = 1;}' is valid anywhere (not just in global scope).  The two 
major differences are the declaration of parsable variables and arrays.

\noindent {\bf Parsable Variables}

\begin{figure}[htp!]
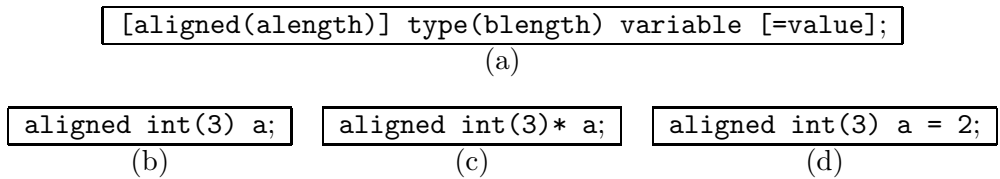

\begin{center}

\begin{tabular}{|c|}                          
\hline
\cf{[aligned(alength)] type(blength) variable [=value]}; \\ 
\hline
\end{tabular} 
\ \\
(a) \\
\ \\

\begin{tabular}{ccc}

\begin{tabular}{|c|}
\hline
\cf{aligned int(3) a}; \\
\hline
\end{tabular} & \begin{tabular}{|c|}
                \hline
                \cf{aligned int(3)* a}; \\
                \hline
                \end{tabular} & \begin{tabular}{|c|}
                                \hline
                                \cf{aligned int(3) a = 2}; \\
                                \hline
                                \end{tabular} \\ 
(b)           & (c)           & (d)

\end{tabular}

\caption{Parsable variable.
         (a) Parsable variable declaration syntax.
         (b) Parsable variable declaration.
         (c) Look-ahead parsing.
         (d) Declaration of a parsable variable with an expected value.}
\label{fig:parse}
\end{center}
\end{figure}

Parsable variables are the core of Flavor's design; it is the proper definition
of these variables that defines the bitstream syntax.  Parsable variables
include a parse length specification immediately after their type declaration,
as shown in Figure~\ref{fig:parse}.  In Figure~\ref{fig:parse}(a), the
{\ct blength} argument can be an integer constant, a non-constant variable of
type compatible to {\ct int}, or a map (discussed later on) with the same type
as the variable.  This means that the parse length of a variable can be
controlled by another variable.  For example, the parsable variable declaration
in Figure~\ref{fig:parse}(b) indicates that the variable {\ct a} has the parse
length of 3 bits.  In addition to the parse length specification, parsable
variables can also have the {\ct aligned} modifier.  This signifies that the
variable begins at the next integer multiple boundary of the length argument -
{\ct alength} -  specified within the alignment expression.  If this length is
omitted, an alignment size of 8 is assumed (byte boundary).  Thus, the variable
{\ct a} is byte-aligned and for parsing, any intermediate bits are ignored, 
while for output bitstream generation the bitstream is padded with zeroes.

As we will see later on, parsable variables cannot be assigned to.  This ensures
that the syntax is preserved regardless if we are performing an input or output
operation.  However, parsable variables {\em can be redeclared}, as long as 
their type remains the same, only the parse size is changed, and the original 
declaration was not as a {\ct const}.  This allows one to select the parse size 
depending on the context (see Expressions and Statements, 
Section~\ref{sec:stmts}).  On top of this, they obey special scoping rules as
described in Section~\ref{sec:scope}.

In general, the parse size expression must be a non-negative value.  The special
value 0 can be used when, depending on the bitstream context, a variable is 
not present in the bitstream but obtains a default value.  In this case, no bits
will be parsed or generated, however, the semantics of the declaration will be 
preserved.  The variables of type {\ct float}, {\ct double}, and {\ct long 
double} are only allowed to have a parse size equal to the fixed size that
their standard representation requires (32 and 64 bits).

\noindent {\bf Look-Ahead Parsing}

In several instances, it is desirable to examine the immediately following
bits in the bitstream without actually removing the bits from the input
stream.  To support this behavior, a `*' character can be placed after the
parse size parentheses.  Note that for bitstream output purposes, this has no
effect. An example of a declaration of a variable for look-ahead parsing is
given in Figure~\ref{fig:parse}(c).

\noindent {\bf Parsable Variables with Expected Values}

Very often, certain parsable variables in the syntax have to have specific 
values (markers, start codes, reserved bits, etc.).  These are specified as 
initialization values for parsable variables.  Figure~\ref{fig:parse}(d) shows 
an example.  The example is interpreted as: {\ct a} is an integer represented
with 3 bits, and must have the value 2. The keyword {\ct const} may be
prepended in the declaration, to indicate that the parsable variable will have
this constant value and, as a result, cannot be redeclared.

As both parse size and initial value can be arbitrary expressions, we should 
note that the order of evaluation is parse expression first, followed by the 
initializing expression.

\noindent {\bf Arrays}

Arrays have special behavior in Flavor, due to its declarative nature but also
due to the desire for very dynamic type declarations.  For example, we want to 
be able to declare a parsable array with different array sizes depending on the
context.  In addition, we may need to load the elements of an array one at a 
time (this is needed when the retrieved value indicates indirectly if further 
elements of the array should be parsed).  These concerns are only relevant for 
parsable variables.  The array size, then, does not have to be a constant 
expression (as in C++ and Java), but it can be a variable as well.  The example 
in Figure~\ref{fig:array}(a) is allowed in Flavor.

\begin{figure}[htp!]
\begin{center}

\begin{tabular}{cccc}

\begin{tabular}{|l|}
\hline
\cf{int(5) a;}     \\
\cf{int(2) A[a]}; \\
\hline
\end{tabular}
    & \begin{tabular}{|l|}
      \hline
      \cf{int A[2] = \{1, 2\}}; \\
      \cf{int A[3] = 5}; \\
      \hline
      \end{tabular}
          & \begin{tabular}{|l|}
            \hline
            \cf{int a = 1};             \\
            \cf{int((a++)) A[(a++)] = (a++)}; \\
            \hline
            \end{tabular}
                & \begin{tabular}{|l|}

                  \hline
                  \cf{int(2) A[[3]] = 1}; \\
                  \cf{int(4) B[[2]][3]};  \\
                  \hline
                  \end{tabular} \\
(a) & (b) & (c) & (d)

\end{tabular}

\caption{Array.
         (a) Declaration with dynamic size specification.
         (b) Declaration with initialization.
         (c) Declaration with dynamic array and parse size.
         (d) Declaration of partial arrays.} 
\label{fig:array}
\end{center}
\end{figure}

An interesting question is how to handle initialization of arrays, or parsable
arrays with expected values.  In addition to the usual brace-expression 
initialization, Flavor also provides a mechanism that involves the specification 
of a single expression as the initializer as shown in Figure~\ref{fig:array}(b). 
This means that all elements of {\ct B} will be initialized with the value 5. 
In order to provide more powerful semantics to array initialization, Flavor 
considers the parse size and initializer expressions as executed per each 
element of the array.  The array size expression, however, is only executed once, 
before the parse size expression or the initializer expression.

Let's look at a more complicated example in Figure~\ref{fig:array}(c).  Here 
{\ct A} is declared as an array of~2 integers.  The first one is parsed with~3 
bits and is expected to have the value~4, while the second is parsed with~5 
bits and is expected to have the value~6.  After the declaration, {\ct a} is 
left with the value~6.  This probably represents the largest deviation of 
Flavor's design from C++ and Java declarations.  On the other hand it does 
provide significant flexibility in constructing sophisticated declarations in a
very compact form, and it is also in line with the dynamic nature of variable 
declarations that Flavor provides.

\noindent {\bf Partial Arrays}

An additional refinement of array declaration is partial arrays.  These are 
declarations of parsable arrays in which only a subset of the array needs to be
declared (or, equivalently, parsed from or written to a bitstream).  Flavor 
introduces a double bracket notation for this purpose.  The example in 
Figure~\ref{fig:array}(d) demonstrates its use.  In the first line, we are 
declaring the~4-th element of {\ct A} (array indices start from~0).  The array 
size is unknown at this point, but of course it will be considered at least 4. 
In the second line, we are declaring a two-dimensional array, and in particular
only its third row (assuming the first index corresponds to a row).  The array 
indices can, of course, be expressions themselves. Partial arrays can only 
appear on the left-hand side of declaration and are not allowed in 
expressions.

\subsection{Expressions and Statements}
\label{sec:stmts}

Flavor supports all of the C++ and Java arithmetic, logical and assignment
operators.  However, parsable variables cannot be used as lvalues.  This ensures
that they always represent the bitstream's content, and allow consistent
operations for the translator-generated {\ct get()} and {\ct put()} methods
that read and write, respectively, data according to the specified form.
Refer to Section~\ref{sec:rtapi} for detailed information about these methods.

Flavor also supports all the familiar flow control statements: {\ct if-else}, 
{\ct do-while}, {\ct while}, and {\ct switch}.  In contrast to C++ and Java, 
variable declarations are not allowed within the arguments of these statements 
(i.e., `{\ct for(int i=0; ; );}' is not allowed).  This is because in C++ the 
scope of this variable will be the enclosing one (even though this is fixed in the 
new ANSI C++ standard, some compilers still use the older concept), while in Java 
it will be the enclosed one.  To avoid confusion, we opted for the exclusion of both 
alternatives at the expense of a slightly more verbose notation. Scoping rules 
are discussed in detail in Section~\ref{sec:scope}.  Similarly, Java only allows
Boolean expressions as part of the flow control statements, and statements like
`{\ct if (1)} \{ ... \}' is not allowed in Java.  Thus, only the flow control
statements with boolean expressions are valid in Flavor.

Figure~\ref{fig:cond} shows an example of the use of these flow control
statements.  The variable {\ct b} is declared with a parse size of 16 if {\ct
a} is equal to 1, and with a parse size of 24 otherwise.  Observe that this
construct would not be meaningful in C++ or Java as the two declarations would
be considered as being in separate scopes.  This is the reason why parsable
variables need to obey slightly different scoping rules than regular
variables.  The way to approach this to avoid confusion is to consider that
Flavor is designed so that these parsable variables have to be properly
defined at the right time and position.  All the rest of the code is there to
ensure that this is the case.  We can consider the parsable variable
declarations as ``actions'' that our system will perform at the specified
times.  This difference, then, in the scoping rules becomes a very natural one.

\begin{figure}[htp!]
\begin{center}

\begin{tabular}{|l|}
\hline
\cf{if (a == 1)} \{                                                                              \\
\hspace{.2in}\cf{little int(16) b}; \cf{// b is a 16 bit integer in little-endian byte ordering} \\
\} \cf{else} \{                                                                                  \\
\hspace{.2in}\cf{little int(24) b}; \cf{// b is a 24 bit integer in little-endian byte ordering} \\
\}                                                                                               \\
\hline
\end{tabular}

\caption{Example of a conditional expression.} 
\label{fig:cond}
\end{center}
\end{figure}


\subsection{Classes}
\label{sec:classes}

Flavor uses the notion of classes in exactly the same way as C++ and Java do. 
It is the fundamental structure in which object data are organized.  Keeping in 
line with the support of both C++ and Java-style programming, classes in Flavor
cannot be nested, and only single inheritance is supported.  In addition, due to
the declarative nature of Flavor, methods are not allowed (this includes 
constructors and destructors as well).

Figure~\ref{fig:class1} shows an example of a simple class declaration with 
just two parsable member variables.  The trailing `;' character is optional 
accommodating both C++ and Java-style class declarations. This class defines 
objects which contain two parsable variables.  They will be present in the 
bitstream in the same order they are declared.  After this class is defined, we 
can declare objects of this type as follows: `{\ct SimpleClass a;}'.

\begin{figure}[htp!]
\begin{center}

\begin{tabular}{|l|}
\hline
\cf{class SimpleClass} \{                 \\
\hspace{.2in}\cf{int(3) a};               \\
\hspace{.2in}\cf{unsigned int(4) b};      \\
\}; \cf{// The trailling `;' is optional} \\
\hline
\end{tabular}                          

\caption{A simple class declaration.} 
\label{fig:class1}
\end{center}
\end{figure}

A class is considered parsable if it contains at least one variable that is 
parsable.  Declaration of parsable class variables can be prepended by the 
{\ct aligned} modifier in the same way as parsable variables.

Class member variables in Flavor do not require access modifiers ({\ct public},
{\ct protected}, or {\ct private}).  In essence, all such variables are considered 
public.

\subsubsection{Parameter Types}
\label{sec:params}

As Flavor classes cannot have constructors, it is necessary to have a mechanism
to pass external information to a class.  This is accomplished using 
{\em parameter types}.  These act the same way as formal arguments in function 
or method declarations do.  They are placed in parentheses after the name of the
class. Figure~\ref{fig:class2} gives an example of a simple class
declaration with parameter types.  When declaring variables of parameter type
classes, it is required that the actual arguments are provided in place of the
formal ones as displayed in the figure.

Of course the types of the formal and actual parameters must match.  For arrays,
only their dimensions are relevant; their actual sizes are not significant as 
they can be dynamically varying.  Note that class types are allowed in parameter
declarations as well.

\begin{figure}[htp!]
\begin{center}

\begin{tabular}{|l|} 
\hline
\cf{class SimpleClass(int i[2])} \{         \\
\hspace{.2in}\cf{int(3) a = i[0]};          \\
\hspace{.2in}\cf{unsigned int(4) b = i[1]}; \\
\};                                         \\ 
\ \\
\cf{int(2) v[2]};      \\
\cf{SimpleClass a(v)}; \\
\hline
\end{tabular} 

\caption{A simple class declaration with parameter types.} 
\label{fig:class2}
\end{center}
\end{figure}

\subsubsection{Inheritance}
\label{sec:inherit}

As we mentioned earlier, Flavor supports single inheritance so that 
compatibility with Java is maintained.  Although Java can ``simulate'' multiple 
inheritance through the use of interfaces, Flavor has no such facility (it 
would be meaningless since methods do not exist in Flavor).  However, for media 
representation purposes, we have not found any instance where multiple 
inheritance would be required, or even be desirable.  It is interesting to note
that all existing representation standards today are not truly object-based. 
The only exception, to our knowledge, is the MPEG-4 specification that 
explicitly addresses the representation of audio-visual objects.  It is, of 
course, possible to describe existing structures in an object-oriented way but
it does not truly map one-to-one with the notion of objects.  For example, 
the MPEG-2 Video slices can be considered as separate objects of the same type,
but of course their semantic interpretation (horizontal stripes of macroblocks)
is not very useful.  Note that containment formats like MP4 (MPEG-4 Systems
file format) and Apple's Quick Time are more object-oriented, as they are 
composed of object-oriented structures called ``atoms''.

Derivation in C++ and Java is accomplished using a different syntax ({\ct extends} 
versus `:'). Here we opted for the Java notation (also `:' is used for object 
identifier declarations as explained below).  Unfortunately, it was not possible 
to satisfy both.

In Figure~\ref{fig:inherit}(a) we show a simple example of a derived class 
declaration.  Derivation from a bitstream representation point of view means 
that {\ct B} is an {\ct A} with some additional information.  In other words, 
the behavior would be almost identical if we just copied the statements between
the braces in the declaration of {\ct A} in the beginning of {\ct B}.  We say 
`almost' here because scoping rules of variable declarations also come into play,
as discussed in Section~\ref{sec:scope}.

Note that if a class is derived from a parsable class, it is also considered 
parsable.

\begin{figure}[htp!]
\begin{center}

\begin{tabular}{cc}
\begin{tabular}{|l|}
\hline
\cf{class A} \{             \\
\hspace{.2in}\cf{int(2) a}; \\
\}                          \\
\                           \\
\cf{class B extends A} \{   \\
\hspace{.2in}\cf{int(3) b}; \\
\}                          \\
\hline
\end{tabular}
    & \begin{tabular}{|l|}
      \hline
      \cf{class A:int(1) id = 0} \{           \\
      \hspace{.2in}\cf{int(2) a};             \\
      \}                                      \\
      \                                       \\
      \cf{class B extends A:int(1) id = 1} \{ \\
      \hspace{.2in}\cf{int(3) b};             \\
      \}                                      \\
      \hline
      \end{tabular} \\ 
(a) & (b) \\
\   & \   \\
\end{tabular}

\begin{tabular}{|l|}
\hline
\cf{class slice : aligned bit(32) slice\_start\_code 
= 0x00000101 .. 0x000001AF} \{                       \\
\hspace{.2in}\cf{\dots}                              \\
\}                                                   \\
\hline
\end{tabular}
\ \\
(c)

\caption{Inheritance.
         (a) Derived class declaration.
         (b) Derived class declaration with object identifiers.
         (c) Class with ID range.} 
\label{fig:inherit}
\end{center}
\end{figure}

\subsubsection{Polymorphic Parsable Classes}
\label{sec:poly}

The concept of inheritance in object-oriented programming derives its power 
from its capability to implement polymorphism.  In other words, the capability 
to use a derived object in a place where an object of the base class is 
expected.  Although the mere structural organization is useful as well, it could
be accomplished equally well with containment (a variable of type {\ct A} is 
the first member of {\ct B}).

Polymorphism in traditional programming languages is made possible via vtable 
structures, which allow the resolution of operations during run-time.  Such 
behavior is not pertinent for Flavor, as methods are not allowed.

A more fundamental issue, however, is that Flavor describes the bitstream 
syntax: the information with which the system can detect which object to select
{\em must be present in the bitstream}.  As a result, traditional inheritance 
as defined in the previous section {\em does not} allow the representation of 
polymorphic objects.  Considering Figure~\ref{fig:inherit}(a), there is no way
to figure out, by reading a bitstream, if we should read an object of type
{\ct A} or type {\ct B}.

Flavor solves this problem by introducing the concept of {\em object 
identifiers} or IDs.  The concept is rather simple: in order to detect which 
object we should parse/generate, there must be a parsable variable that will 
identify it.  This variable must have a different expected value for any class
derived from the originating base class, so that object resolution can be 
uniquely performed in a well-defined way (this can be checked by the translator). 
As a result, object ID values must be constant expressions and they are
always considered constant, i.e., they cannot be redeclared within the class.

In order to signify the importance of the ID variables, they are declared 
immediately after the class name (including any derivation declaration) and 
before the class body.  They are separated from the class name declaration 
using a colon (`:').  We could rewrite the example of Figure~\ref{fig:inherit}(a) 
with IDs as shown in Figure~\ref{fig:inherit}(b).  Upon reading the bitstream, 
if the next 1 bit has the value 0, an object of type {\ct A} will be parsed; 
if the value is 1, then an object of type {\ct B} will be parsed.  For output
purposes, and as will be discussed in Section~\ref{sec:flavort}, it is up to 
the user to set up the right object type in preparation for output.

The name and the type of the ID variable is irrelevant, and can be anything 
that the user chooses.  It cannot, however, be an array or a class variable 
(only built-in types are allowed).  Also, the name, type and parse size must 
be identical between the base and derived classes.  However, object 
identifiers are not required for all derived classes of a base class that has
a declared ID.  In this case, only the derived classes with defined IDs can be
used wherever the base class can appear.  This type of polymorphism is already
used in the MPEG-4 Systems specification, and in particular the Binary Format
for Scenes (BIFS) \cite{mn:mpeg4sys}.  This is a VRML-derived set of nodes 
that represent objects and operations on them, thus forming a hierarchical
description of a scene.

The ID of a class is also possible to have a range of possible values which is 
specified as start\_id \dots end\_id, inclusive of both bounds.  See
Figure~\ref{fig:inherit}(c) for example.

\subsection{Scoping Rules}
\label{sec:scope}

The scoping rules that Flavor uses are identical with C++ and Java with the 
exception of parsable variables.  As in C++ and Java, a new scope is introduced 
with curly braces (\{\}).  Since Flavor does not have functions or methods, a 
scope can either be the global one or a scope within a class declaration.  Note
that the global scope cannot contain any parsable variable, since it does not 
belong to any object.  As a result, global variables can only be constants.

Within a class, all parsable variables are considered as class member 
variables, regardless of the scope they are encountered in.  This is essential 
in order to allow conditional declarations of variables which will almost 
always require that the actual declarations occur within compound statements 
(see Figure~\ref{fig:cond}).  Non-parsable variables that occur in the top-most
class scope are also considered class member variables.  The rest live within 
their individual scopes.

This distinction is important in order to understand which variables are 
accessible to a class variable that is contained in another class.  The issues 
are illustrated in Figure~\ref{fig:scope}.  Looking at the class {\ct A}, the 
initial declaration of {\ct i} occurs in the top-most class scope; as a result 
{\ct i} is a class member.  The variable {\ct a} is declared as a parsable
variable, and hence it is automatically a class member variable.  The
declaration of {\ct j} occurs in the scope enclosed by the {\ct if} statement; 
as this is not the top-level scope, {\ct j} is not a class member.  The
following declaration of {\ct i} is acceptable; the original one is hidden
within that scope.  Finally, the declaration of the variable {\ct a} as a
non-parsable would hide the parsable version.  As parsable variables do not obey
scoping rules, this is not allowed (hiding parsable variables of a base class,
however, is allowed).  Looking now at the declaration of the class {\ct B} which
contains a variable of type {\ct A}, it becomes clear which variables are
available as class members.

\begin{figure}[htp!]
\begin{center}

\begin{tabular}{|cc|}
\hline

\begin{tabular}{l}
\cf{class A} \{                                           \\
\hspace{.2in}\cf{int i = 1};                              \\
\hspace{.2in}\cf{int(2) a};                               \\
\hspace{.2in}\cf{if (a == 2)} \{                          \\
\hspace{.4in}\cf{int j = i};                              \\
\hspace{.4in}\cf{int i = 2; // Hides i, ok}               \\
\hspace{.4in}\cf{int a; // Hides a, error}                \\
\hspace{.2in}\}                                           \\
\}
\end{tabular} &

     \begin{tabular}{l}
     \cf{class B} \{                                           \\
     \hspace{.2in}\cf{A a};                                    \\
     \hspace{.2in}\cf{a.j = 1; // Error, j not a class member} \\
     \hspace{.2in}\cf{int j = a.a + 1; // Ok}                  \\
     \hspace{.2in}\cf{j = a.i + 2 // Ok}                       \\
     \hspace{.2in}\cf{int(3) b};                               \\
     \}                                                        \\
     \                                                         \\
     \       
     \end{tabular} \\

\hline
\end{tabular}

\caption{Scoping rules example.} 
\label{fig:scope}
\end{center}
\end{figure}

In summary, the scoping rules have the following two special considerations. 
Parsable variables do not obey scoping rules and are always considered class 
members.  Non-parsable variables obey the standard scoping rules and are 
considered class members only if they are at the top-level scope of the class.

Note that parameter type variables are considered as having the top-level 
scope of the class.  Also, they are not allowed to hide the object identifier, 
if any.

\subsection{Maps}
\label{sec:maps}

Up to now, we have only considered fixed-length representations, either 
constant or parametric.  A wide variety of representation schemes, however, 
rely heavily on entropy coding, and in particular Huffman codes
~\cite{bk:cover:info}.  These are variable-length codes (VLCs) which 
are uniquely decodable (no codeword is the prefix of another).  Flavor provides 
extensive support for variable-length coding through the use of maps.  These 
are declarations of tables in which the correspondence between codewords and 
values is described.

Figure~\ref{fig:map}(a) gives a simple example of a map declaration.  The
{\ct map} keyword indicates the declaration of a map named {\ct A}.  The
declaration also indicates that the map converts from bitstring values to 
values of type {\ct int}.  The type indication can be a fundamental type, a 
class type, or an array.  Map declarations can only occur in global scope.  As a 
result, an array declaration will have to have a constant size (no non-constant
variables are visible at this level).  After the map is properly declared, we 
can define parsable variables that use it by indicating the name of the map
where we would put the parse size expression as follows: {\ct int(A) i}.  As we
can see the use of variable-length codes is essentially identical to
fixed-length variables.  All the details are hidden away in the map declaration.

\begin{figure}[htp!]
\begin{center}

\begin{tabular}{cc}

\begin{tabular}{c}
\begin{tabular}{|l|}
\hline
\cf{map A(int)} \{        \\
\hspace{.2in}\cf{0b0, 1,} \\
\hspace{.2in}\cf{0b01, 2} \\
\}                        \\
\hline
\end{tabular} \\
(a)           \\
\             \\
\begin{tabular}{|l|}
\hline
\cf{map A(int)} \{              \\
\hspace{.2in}\cf{0b0, 1,}       \\
\hspace{.2in}\cf{0b01, 2,}      \\
\hspace{.2in}\cf{0b001, int(5)} \\
\}                              \\
\hline
\end{tabular} \\
(c)           
\end{tabular} 
    & \begin{tabular}{c}
      \begin{tabular}{|l|}
      \hline
      \cf{// The output type of a map is defined in a class} \\
      \cf{class YUVblocks} \{                                \\
      \hspace{.2in}\cf{unsigned int Yblocks};                \\
      \hspace{.2in}\cf{unsigned int Ublocks};                \\
      \hspace{.2in}\cf{unsigned int Vblocks};                \\
      \}                                                     \\
      \                                                      \\
      \cf{// A table that relates the chroma format with}    \\
      \cf{// the number of blocks per signal component}      \\
      \cf{map blocks\_per\_component (YUVblocks)} \{         \\
      \hspace{.2in}\cf{0b00,} \{{\cf 4, 1, 1}\}{\cf ,   // 4:2:0}  \\
      \hspace{.2in}\cf{0b01,} \{{\cf 4, 2, 2}\}{\cf ,   // 4:2:2}  \\
      \hspace{.2in}\cf{0b10,} \{{\cf 4, 4, 4}\}{\cf \ \ // 4:4:4}  \\
      \}                                                     \\
      \hline
      \end{tabular} \\
      (b)
      \end{tabular}

\end{tabular}

\caption{Map.
         (a) A simple map declaration.
         (b) A map with defined output type.
         (c) A map declaration with extension.}  
\label{fig:map}
\end{center}
\end{figure}

The map contains a series of entries.  Each entry starts with a bit string that 
declares the codeword of the entry followed by the value to be assigned to 
this codeword.  If a complex type is used for the mapped value, then the values 
have to be enclosed in curly braces.  Figure~\ref{fig:map}(b) shows the 
definition of a VLC table with a user-defined class as output type.  The type of
the variable has to be identical to the type returned from the map.  For 
example, using the declaration - {\ct YUVblocks(blocks\_per\_component) 
chroma\_format;} - we can access a particular value of the map using the
construct: {\ct chroma\_format.Ublocks}.

As Huffman codeword lengths tend to get very large when their number increases,
it is typical to specify ``escape codes,'' signifying that the actual value
will be subsequently represented using a fixed-length code.  To accommodate
these as well as more sophisticated constructs, Flavor allows the use of
parsable type indications in map values.  This means that, using the example in 
Figure~\ref{fig:map}(a), we can write the example in Figure~\ref{fig:map}(c).
This indicates that, when the bit string {\ct 0b001} is encountered in the 
bitstream, the actual return value for the map will be obtained by parsing 5 
more bits.  The parse size for the extension can itself be a map, thus allowing 
the cascading of maps in sophisticated ways.  Although this facility is 
efficient when parsing, the bitstream generation operation can be costly when 
complex map structures are designed this way.  None of today's specifications 
that we are aware of require anything beyond a single escape code.

The translator can check that the VLC table is uniquely decodable, and also
generate optimized tables for extremely fast encoding/decoding using our
hybrid approach as described in Section~\ref{sec:mapproc}.

\subsection{Built-In Operators}
\label{sec:operators}

Operators are built-in functions that are made available to the Flavor programmer
in order to facilitate certain frequently appearing data manipulations.  These 
operators are the only functions that are available in the current version of 
Flavor.

\subsubsection{lengthof()}

The {\ct lengthof()} operator is used to obtain the parsed length of a parsable
variable.  The general syntax is as follows: `{\ct lengthof(var);}' where {\ct
var} must be a parsable variable of any type that has been previously declared.
The result of the operator is treated as an integer.

Since parsable variables can be declared more than once, the operator considers
only the last instance of the variable that has been declared (parsed from the 
bitstream).  The following (Figure~\ref{fig:lengthof}) is an example of the use 
of the operator.

\begin{figure}[htp!]
\begin{center}

\begin{tabular}{cc}

\begin{tabular}{|l|}
\hline
\cf{int(5) i = 3;}           \\
\cf{...}                     \\
\cf{int(3) j = lengthof(i);} \\
\hline
\end{tabular} 
    & \begin{tabular}{|l|}
      \hline
      \cf{int i = 1;}              \\
      \cf{int(i++) a[5];}          \\
      \cf{int j = lengthof(a[4]);} \\
      \hline
      \end{tabular} \\
(a) & (b)

\end{tabular}

\caption{{\ct lengthof()}.
         (a) A simple example.
         (b) A more complex example where a multi-dimensional array is used.}  
\label{fig:lengthof}
\end{center}
\end{figure}

In Figure~\ref{fig:lengthof}(b), we declare an one-dimensional array {\ct a}
with five elements.  The variable {\ct j} is set with the length of the last 
(4-th) element.  It is easy to see that the length of this element is going
to be 5.

In a declaration such as `{\ct aligned class A\{\}}', the bits skipped for
alignment are not accounted for by the {\ct lengthof()} operator.  This is
true for simple variables as well.  These bits, however, are counted for the 
enclosing class.

\subsubsection{isidof()}

The {\ct isidof()} operator is used to check if the value of a varible is 
among the IDs of a polymorphic class.  The general syntax is as follows:
`{\ct isidof(class\_name, var);}' where {\ct class\_name} is the name of a 
polymorphic class that has been previously declared, and {\ct var} is the name 
of a simple variable.  The result of the operator is treated as an integer, 
and it is 1 if the value of {\ct var} is among the IDs of {\ct class\_name} 
or 0, otherwise.

This operator was introduced to accommodate a coding structure in which the 
syntax was expressed as: pase as many objects of a particular type as you can 
from the bitstream.  This corresponds to examining the following bits on the
bitstream and, if they correspond to an object of the given type, parsing it;
otherwise the syntax would continue to the next constrcut.  Without this 
operator, the programmer would have to explicitly construct a {\ct switch}
statement or a series of {\ct if}-{\ct then}-{\ct else} statements, checking
against all IDs of the class.  This not only would be tiresome, but would 
also be a source of errors if one of the IDs were not included.

The following is an example of the use of this operator for parsing the 
above-mentioned constructs.

\begin{figure}[htp!]
\begin{center}

\begin{tabular}{|l|}
\hline
\cf{abstract class A : int(8) id = 0 \{} \\
\cf{...}                                 \\
\cf{\}}                                  \\
\cf{...}                                 \\
\cf{int(8)* id;}                         \\
\cf{int i;}                              \\
\cf{while (isidof(A, id)==1) \{}         \\
\hspace{.2in}\cf{A a[[i++]];}            \\
\hspace{.2in}\cf{int(8)* id;}            \\
\cf{\}}                                  \\
\hline
\end{tabular} 

\caption{An example of using the {\ct isidof()} operator.}  
\label{fig:isidof}
\end{center}
\end{figure}

In Figure~\ref{fig:isidof}, we declare an (abstract) class {\ct A}, with 
presumably a number of derived classes (not show here).  The {\ct while} loop
following the declaration of {\ct A} examines the next 8 bits of the bitstream.
If they correspond to an ID of one of the classes derived from {\ct A} then the 
object is parsed; if not, the code continues.

\section{The Flavor Translator}
\label{sec:flavort}

Designing a language like Flavor would be an interesting but academic exercise,
unless it was accompanied by software that can put its power into full use. We 
have developed a translator that evolved concurrently with the design of the 
language. When the language specification became stable, the translator was 
completely rewritten. The most recent release is publicly available for 
downloading at {\ct http://www.sourceforge.net/projects/flavor}.

\subsection{Run-Time API}
\label{sec:rtapi}

The translator reads a Flavor source file ({\ct .fl}) and, depending on the
language selection flag, it generates a pair of {\ct .h} and 
{\ct .cpp} files (for C++) or a set of {\ct .java} files (for Java).  In
the case of C++, the {\ct .h} file contains the declarations of all Flavor
classes as regular C++ classes and the {\ct .cpp} file contains the
implementations of the corresponding class methods ({\ct put()} and {\ct
get()}). In the case of Java, each {\ct .java} file contains the declaration
and implementation of a single Flavor class. In both cases, the {\ct get()}
method is responsible for reading a bitstream and loading the class variables
with their appropriate values, while the {\ct put()} method does the
reverse. All the members of the classes are declared {\ct public}, and this 
allows direct access to desired fields in the bitstream.

The translator makes minimal assumptions about the operating environment for
the generated code. For example, it is impossible to anticipate all possible
I/O structures that might be needed by applications (network-based, 
multi-threaded, multiple buffers, etc.). Attempting to provide a universal 
solution would be futile. Thus, instead of having the translator directly output 
the required code for bitstream I/O, error reporting, tracing, etc., a run-time
library is provided. With this separation, programmers have the flexibility of 
replacing parts of, or the entire library with their own code. The only 
requirement is that the customized code provides an identical interface to the
one needed by the translator. This interface is defined in a pure virtual class 
called {\ct IBitstream}. Deriving from the provided {\ct IBitstream} class will
ensure compatibility with the translator. Additionally, as the source code for 
the library is included in its entirety, customization can be performed quite 
easily. The Flavor package also provides information on how to rebuild the 
library, if needed.

The run-time library includes the {\ct Bitstream} class that is derived from the 
{\ct IBitstream} interface, and provides basic bitstream I/O facilities in terms 
of reading or writing bits from a binary file. A {\ct Bitstream} reference is 
passed as an argument to the {\ct get()} and {\ct put()} methods. 

If parameter types are used in a class, then they are also required arguments 
in the {\ct get()} and {\ct put()} methods as well. The translator also 
requires that a function is available to receive calls when expected values are
not available or VLC lookups fail. The function name can be selected by the 
user; a default implementation ({\ct flerror}) is included in the run-time
library.

For efficiency reasons, Flavor arrays are converted to fixed size arrays in the
translated code. This is necessary in order to allow developers to access 
Flavor arrays without needing special techniques. Whenever possible, the 
translator automatically detects and sets the maximum array size; it can also 
be set by the user using a command-line option. Finally, the run-time library 
(and the translator) only allows parse sizes of up to the native integer size 
of the host processor (except for double values). This enables fast 
implementation of bitstream I/O operations.

For parsing operations, the only task required by the programmer is to declare 
an object of the class type at hand, and then call its {\ct get()} method with 
an appropriate bitstream. While the same is also true for {\ct put()} 
operation, the application developer must also load all class member variables
with their appropriate values before the call is made.

\subsection{Include and Import Directives}
\label{sec:directives}

In order to simplify the source code organization, Flavor supports 
{\ct \%include} and {\ct \%import} directives. These are the mechanisms to
combine several different source code files into one entity, or to share a
given data structure definition across different projects.

\subsubsection{Include Directive}
\label{sec:include}

The statement - {\ct \%include "file.fl"} - will include the specified 
{\ct .fl} file in the current position and will flag all of its content so that
no code is generated. Figure~\ref{fig:directive1}(a) displays a {\ct .fl} file
({\ct other.fl}) that is included by another {\ct .fl} file ({\ct main.fl}).
The {\ct other.fl} file contains the definition of the constant {\ct a}. The
inclusion makes the declaration of the {\ct a} variable available to the
{\ct main.fl} file. In terms of the generated output,
Figure~\ref{fig:directive1}(b) outlines the placement of information in 
different files. In the figure, we see that the main and included files each 
keep their corresponding implementations. The generated C++ code maintains this
partitioning, and makes sure that the main file includes the C++ header file of
the included Flavor file.

The {\ct \%include} directive is useful when data structures need to be shared
across modules or projects. It is similar in spirit to the use of the C/C++
preprocessor {\ct \#include} statement in the sense that it is used to make
general information available at several different places in a program. Its
operation, however, is different as Flavor's {\ct \%include} statement does not
involve code generation for the included code. In C/C++, {\ct \#include} is 
equivalent to copying the included file in the position of the {\ct \#include}
statement. This behavior is offered in Flavor by the {\ct \%import} directive.

Similarly, when generating the Java code, only the {\ct .java} files 
corresponding to the currently processed Flavor file are generated. The
data in the included files are allowed to be used, but they are not generated.

\begin{figure}[htp!]
\begin{center}

\begin{tabular}{|l|}
\hline
\cf{// In the file, other.fl}                                                      \\
\cf{const int a = 4};                                                              \\
\                                                                                  \\
\cf{// In the file, main.fl}                                                       \\
\cf{\%include "other.fl"}                                                          \\
\                                                                                  \\
\cf{class Test} \{                                                                 \\
\hspace{.2in}\cf{int(a) t; // The variable 'a' is included from the other.fl file} \\
\}                                                                                 \\
\hline
\end{tabular}
\ \\ 
(a) \\
\ \\

\begin{tabular}{|ccc|}
\hline
\begin{tabular}{l}
\cf{// In the file, other.h} \\
\cf{extern const int a};     \\
\ 
\end{tabular}
    & \begin{tabular}{l}
      \cf{// In the file, other.cpp} \\
      \cf{\#include "other.h"}       \\        
      \cf{const int a = 4};         
      \end{tabular}
          & \begin{tabular}{l}
            \cf{// In the file, main.h} \\
            \cf{\#include "other.h"}    \\
            ...                       
            \end{tabular} \\
\hline
\end{tabular}
\ \\
(b)

\caption{The {\ct \%include} directive.
         (a) The {\ct other.fl} file is included by the {\ct main.fl} file.
         (b) The {\ct other.h} and {\ct other.cpp} files are generated from the 
             {\ct other.fl} file whereas the {\ct main.h} file is generated from the 
             {\ct main.fl} file.}
\label{fig:directive1}
\end{center}
\end{figure}

\subsubsection{Import Directive}
\label{sec:import}

The {\ct \%import} directive behaves similarly to the {\ct \%include}
directive, except that full code is generated for the imported file by the 
translator, and no C++ {\ct \#include} statement is used. This behavior is
identical to how a C++ preprocesor {\ct \#include} statement would behave in
Flavor. 

\begin{figure}[htp!]
\begin{center}

\begin{tabular}{|l|}
\hline
\cf{\%import "other.fl"}                                                           \\
\                                                                                  \\
\cf{class Test} \{                                                                 \\
\hspace{.2in}\cf{int(a) t; // The variable 'a' is included from the other.fl file} \\
\}                                                                                 \\
\hline
\end{tabular}
\ \\
(a) \\
\ \\

\begin{tabular}{|cc|}
\hline
\begin{tabular}{l}
\cf{// In the file, main.h} \\
\cf{extern const int a};    \\
...                         
\end{tabular}
    & \begin{tabular}{l}
      \cf{// In the file, main.cpp} \\
      \cf{const int a = 4};         \\
      ...                          
      \end{tabular} \\
\hline
\end{tabular}
\ \\
(b)

\caption{The {\ct \%import} directive.
         (a) The {\ct main.fl} file using the {\ct \%import} directive.
         (b) The {\ct main.h} and {\ct main.cpp} files generated from the {\ct main.fl} 
             file defined in (a).}
\label{fig:directive2}
\end{center}
\end{figure}

Let's consider the example of the previous section, this time with an 
{\ct \%import} directive rather than an {\ct \%include} one as shown in 
Figure~\ref{fig:directive2}(a). As can be seen from 
Figure~\ref{fig:directive2}(b), the generated code includes the C++ code
corresponding to the imported {\ct .fl} file. Therefore, using the
{\ct \%import} directive is exactly the same as just copying the code in the 
imported {\ct .fl} file and pasting it in the same location as the 
{\ct \%import} statement is specified. The translator generates the Java code
in the same way.  

Note that the Java import statement behaves more like Flavor's {\ct \%include}
statement, in that no code generation takes place for the imported (included)
code.

\subsection{Pragma Statements}
\label{sec:pragmas}

Pragma statements are used as a mechanism for setting translator options from 
inside a Flavor source file. This allows modification of translation parameters
(set by the command-line options) without modifying the makefile that builds
the user's program, but more importantly, it allows very fine control on which
translation options are applied to each class, or even variable. Almost all
command-line options have pragma equivalents. The ones excluded were not
considered useful for specification within a source file. 

Pragma statements are introduced with the {\ct \%pragma} directive. It can
appear wherever a statement or declaration can. It can contain one or more
settings, separated by commas, and it cannot span more than one line. After a
setting is provided, it will be used for the remainder of the Flavor file,
unless overridden by a different pragma setting. In other words, pragma 
statements do not follow the scope of Flavor code. A pragma that is included
in a class will affect not only the class where it is contained, but all
classes declared after it. An example is provided in Figure~\ref{fig:pragma}.

\begin{figure}[htp!]
\begin{center}

\begin{tabular}{|l|}
\hline
\cf{// Activate both put and get, generate tracing code, and set array size to 128} \\
\cf{\%pragma put, get, trace, array=128}                                            \\
\                                                                                   \\
\cf{class Example} \{                                                               \\
\hspace{.2in}\cf{\%pragma noput // No put() method needed}                          \\
\                                                                                   \\
\hspace{.2in}\cf{unsigned int(10) length};                                          \\
\hspace{.2in}\cf{\%pragma array=1024 // Switch array size to 1024}                  \\
\hspace{.2in}\cf{char(3) data[length]};                                             \\
\                                                                                   \\
\hspace{.2in}\cf{\%pragma array=128 // Switch array size back to 128}               \\
\hspace{.2in}\cf{\%pragma trace="Tracer.trace" // Use custom tracer}                \\
\}                                                                                  \\
\                                                                                   \\
\cf{// The above settings are still active here!}                                   \\
\hline
\end{tabular}

\caption{Some examples of using pragma statements to set the translator options at 
         specific locations.}
\label{fig:pragma}
\end{center}
\end{figure}

In this example, we start off setting the generation of both {\ct get()} and
{\ct put()} methods, enabling tracing and setting the maximum array size to 128
elements. Inside the {\ct Example} class, we disable the {\ct put()} method
output. This class reads a chunk of data, which is preceded by its size
({\ct length}, a 10-bit quantity). This means that the largest possible buffer
size is 1024 elements. Hence for the {\ct data} array that immediately follows,
we set the array size to 1024, and then switch it back to the default of 128. 
Finally, at the end of the class we select a different tracing function name; 
this function is really a method of a class, but this is irrelevant for the
translator. Since this directive is used when the {\ct get()} method code is 
produced, it will affect the entire class despite the fact that it is declared
at its end. 

Note that these pragma settings remain in effect even after the end of the
{\ct Example} class.

\subsection{Verbatim Code} 
\label{sec:verb}

In order to further facilitate integration of Flavor code with C++/Java user 
code, the translator supports the notion of verbatim code. Using special 
delimiters, code segments can be inserted in the Flavor source code, and copied
verbatim at the correct places in the generated C++/Java file. This allows, for
example, the declaration of constructors/destructors, user-specified methods, 
pointer member variables for C++, etc. Such verbatim code can appear wherever 
a Flavor statement or declaration is allowed.

\begin{figure}[htp!]
\begin{center}
\begin{tabular}{|l|}
\hline
\cf{class GIF87a} \{                                                      \\
\hspace{.2in}\cf{char(8) GIFsignature[6] = "GIF87a"; // GIF signature}    \\
\hspace{.2in}\                                                            \\ 
\hspace{.2in}\cf{\%g}\{ \cf{print()}; \cf{\%g}\}                          \\
\hspace{.2in}\                                                            \\
\hspace{.2in}\cf{ScreenDescriptor sd; // A screen descriptor}             \\
\hspace{.2in}\                                                            \\
\hspace{.2in}\cf{// One or more image descriptors}                        \\
\hspace{.2in}\cf{do} \{                                                   \\
\hspace{.4in}\cf{unsigned int(8) end};                                    \\
\hspace{.4in}\                                                            \\
\hspace{.4in}\cf{if (end == ',')} \{ \cf{// We found an image descriptor} \\
\hspace{.6in}\cf{ImageDescriptor id};                                     \\
\hspace{.4in}\}                                                           \\
\hspace{.4in}\cf{if (end == '!')} \{ \cf{// We found an extension block}  \\
\hspace{.6in}\cf{ExtensionBolck eb};                                      \\
\hspace{.4in}\}                                                           \\
\hspace{.4in}\cf{// Everything else is ignored}                           \\
\hspace{.2in}\} \cf{while (end != ';'); // ';' is the end-of-data marker} \\
\hspace{.2in}\                                                            \\
\hspace{.2in}\cf{\%.c}\{                                                  \\
\hspace{.2in}\cf{void print()} \{ \cf{...} \}                             \\
\hspace{.2in}\cf{\%.c}\}                                                  \\
\hspace{.2in}\cf{\%.j}\{                                                  \\
\hspace{.2in}\cf{void print()} \{ \cf{...} \}                             \\
\hspace{.2in}\cf{\%.j}\}                                                  \\
\}                                                                        \\
\hline
\end{tabular}
\caption{A simple Flavor example: the GIF87a header. The usage of verbatim code
         is illustrated.}
\label{fig:verb}
\end{center}
\end{figure} 

The delimiters {\ct \%}\{ and {\ct \%}\} can be used to introduce code that 
should go to the class declaration itself (or the global scope). The delimiters
{\ct \%p}\{ and {\ct \%p}\}, and {\ct \%g}\{ and {\ct \%g}\} can be used to 
place code at exactly the same position they appear in the {\ct put()} and 
{\ct get()} methods, respectively. Finally, the delimiters {\ct \%*}\{ and 
{\ct \%*}\} can be used to place code in both {\ct put()} and {\ct get()} 
methods. To place code specific to C++ or Java, {\ct .c} or {\ct .j} can be
placed before the braces in the delimiters, respectively.  For example, a 
verbatim code to be placed in the {\ct get()} method of the Java code will be 
delimited with {\ct \%g.j}\{ and {\ct \%g.j}\}.

The Flavor package includes several samples on how to integrate user code 
with Flavor-generated code, including a simple GIF parser. 
Figure~\ref{fig:verb} shows a simple example that reads the header of a GIF87a 
file and prints its values. The print statement which prints the values of the
various elements is inserted as verbatim code in the syntax (within 
{\ct \%g}\{ and {\ct \%g}\} markers, since the code should go in the 
{\ct get()} method). The implementation of the print method for C++ code is 
declared within {\ct \%.c}\{ and {\ct \%.c}\}, and for Java, the corresponding
implemenation is defined within {\ct \%.j}\{ and {\ct \%.j}\}. The complete
sample code can be found in the Flavor package.

\subsection{Tracing Code Generation}
\label{sec:trace}

We also included the option to generate bitstream tracing code within the {\ct
get()} method. This allows one to very quickly examine the contents of a
bitstream for development and/or debugging purposes by creating a dump of the
bitstream's content. With this option, and given the syntax of a bitstream
described in Flavor, the translator will automatically generate a complete
C++/Java program that can verify if a given bitstream complies with that
syntax or not. This can be extremely useful for codec development as well as
compliance testing.

\subsection{Map Processing}
\label{sec:mapproc}

Map processing is one of the most useful features of Flavor, as hand-coding 
VLC tables is tedious and error prone. Especially during the development phase
of a representation format, when such tables are still under design, full
optimization within each design iteration is usually not performed. By using 
the translator, such optimization is performed at zero cost. Also, note that 
maps can be used for fixed-length code mappings just by making all codewords
have the same length. As a result, one can very easily switch between fixed and
variable-length mappings when designing a new representation format.

When processing a map, the translator first checks that it is uniquely 
decodable, i.e., no codeword is the prefix of another. It then constructs 
a class declaration for that map, which exposes two methods: {\ct getvlc()} and
{\ct putvlc()}. These take as arguments a bitstream reference as well as a 
pointer to the return type of the map. The {\ct getvlc()} method is reponsible 
for decoding a map entry and returning the decoded value, while the 
{\ct putvlc()} method is responsible for the output of the correct codeword.
Note that the defined class does not perform any direct bitstream I/O itself,
but uses the services of the {\ct Bitstream} class instead. This ensures that
a user-supplied bitstream I/O library will be seamlessly used for map
processing.

According to Fogg's survey on software and hardware VLC architectures 
\cite{ipr:fogg:vlcsurvey}, fast software decoding methods usually exploit 
a variable-size look-ahead window (multi-bit lookup) with lookup tables. For 
optimization, the look-ahead size and corresponding tables can be customized 
for position dependency in the bitstream. For example, for MPEG video, the 
look-ahead can be selected based on the picture type (I, P, or B).

One of the fastest decoding methods is comprised of one huge lookup table 
where every codeword represents an index of the table pointing to the 
corresponding value. However, this costs too much memory. On the other end 
of the spectrum, one of the most memory efficient algorithms would be of the 
form of a binary tree. The tree is traversed one bit at a time, and at each 
stage one examines if a leaf node is reached; if not, the left or right 
branch is taken for the input bit value of 0 or 1, respectively. Though 
efficient in memory, this algorithm is extremely slow, requiring 
{\ct N} (the bit length of the longest codeword) stages of bitstream input 
and lookup.

In \cite{ipr:fang:map}, we adopted a hybrid approach that maintains the space
efficiency of binary tree decoding, and most of the speed associated with 
lookup tables. In particular, instead of using lookup tables, we use 
hierarchical, nested switch statements. Each time the read-ahead size is 
determined by the maximum of a fixed step size and the size of the next 
shortest code. The fixed step size is used to avoid degeneration of the 
algorithm into binary tree decoding. The benefit of this approach is that 
only complete matches require case statements, while all partial matches 
can be grouped into a single default statement (that, in turn, introduces 
another switch statement).

With the above-mentioned approach, the space requirement consists of storage 
of the case values and the comparison code generated by the translator (this 
code consists of just 2 instructions on typical CISC systems, e.g., a Pentium). 
While slightly larger than a simple binary tree decoder, this overhead still 
grows linearly with the number of code entries (rather than exponentially with 
their length). This is further facilitated by the selection of the step size. 
When the incremental code size is small, multiple case statements may be 
assigned to the same codeword, thus increasing the space requirements.

In \cite{ipr:fang:map}, we compared the performance of various techniques, 
including binary tree parsing, fixed step full lookups with different step sizes 
and our hybrid switch statement approach. In terms of time, our technique is 
faster than a hierarchical full-lookup approach with identical step sizes. This 
is because switching consumes little time compared to fixed-step's function 
lookups. Furthermore, it is optimized by ordering the case statements in terms of 
the length of their codeword. As shorter lengths correspond to higher 
probabilities, this minimizes the average number of comparisons per codeword. 

With Flavor, developers can be assured of extremely fast decoding with minimal 
memory requirement due to the optimized code generation. In addition, the development 
effort and time in creating software to process VLCs is nearly eliminated.

\section{Concluding Remarks}

Flavor's design was motivated by our belief that content creation, access, 
manipulation and distribution will become increasingly important for end-users
and developers alike.  New media representation forms will continue to be 
developed, providing richer features and more functionalities for end-users. In 
order to facilitate this process, it is essential to bring syntactic description
on par with modern software development practices and tools.  Flavor can provide
significant benefits in the area of media representation and multimedia 
application development at several levels.

First, it can be used as a media representation documenting tool, substituting 
ad-hoc ways of describing a bitstream's syntax with a well-defined and concise
language.  This by itself is a substantial advantage for defining specifications,
as a considerable amount of time is spent to ensure that such specifications are
unambiguous and bug-free.

Second, a formal media representation language immediately leads to the 
capability of automatically generating software tools, ranging from bitstream 
generators and verifiers, as well as a substantial portion of an encoder or
decoder.

Third, it allows immediate access to the content by any application developer, 
for such diverse use as editing, searching, indexing, filtering, etc.

With appropriate translation software, and a bitstream representation written 
in Flavor, obtaining access to such content is as simple as cutting and 
pasting the Flavor code from the specification into an ASCII file, and running 
the translator.

Flavor, however, does not provide facilities to specify how full decoding of 
data will be performed as it only addresses bitstream syntax description.  For 
example, while the data contained in a GIF file can be fully described by 
Flavor, obtaining the value of a particular pixel requires the addition of LZW 
decoding code that must be provided by the programmer. In several instances, 
such access is not necessary.  For example, a number of tools have been 
developed to do automatic indexing, searching and retrieval of visual content 
directly in the compressed domain for JPEG and MPEG content (see 
\cite{ipr:smith:visualseek,ar:smoliar:indexing}). Such 
tools only require parsing of the coded data so that DCT coefficients are 
available, but do not require full decoding. Also, new techniques, such 
as MPEG-7, will provide a wealth of information about the content without 
the need to decode it.  In all these cases, parsing of the compressed 
information may be the only need for the application at hand.

Finally, Flavor can also be used to redefine the syntax of content in both 
forward and backward compatible ways.  The separation of parsing from the 
remaining code/decoding operations allows its complete substitution as long as 
the interface (the semantics of the previously defined parsable variables) 
remains the same.  Old decoding code will simply ignore the new variables, while 
newly written encoders and decoders will be able to use them.  Use of Java in 
this repect is very useful; its capability to download new class definitions 
opens the door for such downloadable content descriptions that can accompany 
the content itself (similar to self-extracting archives).  This can eliminate 
the rigidity of current standards, where even a slight modification of the 
syntax to accommodate new techniques or functionalities render the content 
useless in non-flexible but nevertheless compliant decoders.

The authors gratefully acknowledge Olivier Avaro (France Telecom), Carsten
Herpel (Thomson Multimedia), and Jean-Claude Dufourd (ENST) for their
contributions in the Flavor specification during the development of the MPEG-4
standard.  We would also like to acknowledge Yihan Fang, who implemented early
versions of the Flavor translator.


\bibliography{flavor}
\bibliographystyle{unsrt}

\end{document}